\documentstyle[11pt]{article}
\newdimen\linewidth
\linewidth=\hsize

\def\mud#1{\hfil $\displaystyle{#1}$\hfil}
\def\rig#1{\hfil $\displaystyle{#1}$}

\def\vcalignn#1{\noindent{\small\hbox to \linewidth{%
  \hss\valign{\vfil\hbox{##}\vfil\cr #1}\hss}}}

\def\lowerhalf#1{\hbox{\raise -0.5\baselineskip\hbox{#1}}}

\def\lbottom{\mathord{\perp}}

\def\bot{\mathord{\lbottom}}

\def\goesto{\hbox to 0pt{\hss $\Rightarrow$ \hss}}

\newbox\tempa
\newbox\tempb
\newdimen\tempc
\newbox\tempd

\def\inruleanhelp#1#2#3{\setbox\tempa=\hbox{$\displaystyle{\mathstrut #2}$}%
                        \setbox\tempb=\vbox{\halign{##\cr
        \mud{#1}\cr
        \noalign{\vskip \the\lineskip}%
        \noalign{\vskip 1pt}%
        \noalign{\hrule height 0pt}%
        \rig{\vbox to 0pt{\vss\hbox to 0pt{${\ #3}$\hss}\vss}}\cr
        \noalign{\hrule}%
        \noalign{\vskip \the\lineskip}%
        \mud{\copy\tempa}\cr}}%
                      \tempc=\wd\tempb
                      \advance\tempc by \wd\tempa
                      \divide\tempc by 2 }

\def\inrulemhelp#1#2#3{\setbox\tempa=\hbox{$\displaystyle{\mathstrut #2}$}%
                        \setbox\tempb=\vbox{\halign{##\cr
        \mud{#1}\cr
        \noalign{\vskip\the\lineskip}%
        \noalign{\hrule}%
        \noalign{\vskip\the\lineskip}%
        \noalign{\vskip 1pt}%
        \noalign{\hrule height 0pt}%
        \rig{\vbox to 0pt{\vss\hbox to 0pt{${\ #3}$\hss}\vss}}\cr
        \noalign{\hrule}%
        \noalign{\vskip\the\lineskip}%
        \mud{\copy\tempa}\cr}}%
                      \tempc=\wd\tempb
                      \advance\tempc by \wd\tempa
                      \divide\tempc by 2 }

\def\inrulemchelp#1#2#3{\setbox\tempa=\hbox{$\displaystyle{\mathstrut #2}$}%
                        \setbox\tempd=\hbox{$\ #3$}%
                        \setbox\tempb=\vbox{\halign{##\cr
        \mud{#1}\cr
        \noalign{\vskip\the\lineskip}%
        \noalign{\vskip 1pt}%
        \noalign{\hrule}%
        \noalign{\vskip\the\lineskip}%
        \noalign{\hrule height 0pt}%
        \rig{\vbox to 0pt{\vss\hbox to 0pt{${\ #3}$\hss}\vss}}\cr
        \noalign{\hrule}%
        \noalign{\vskip\the\lineskip}%
        \mud{\copy\tempa}\cr}}%
                      \tempc=\wd\tempb
                      \advance\tempc by \wd\tempa
                      \divide\tempc by 2 }

\def\inruleanchelp#1#2#3{\setbox\tempa=\hbox{$\displaystyle{\mathstrut #2}$}%
                        \setbox\tempd=\hbox{$\ #3$}%
                        \setbox\tempb=\vbox{\halign{##\cr
        \mud{#1}\cr
        \noalign{\vskip\the\lineskip}%
        \noalign{\vskip 1pt}%
        \noalign{\hrule height 0pt}%
        \rig{\vbox to 0pt{\vss\hbox to 0pt{\copy\tempd \hss}\vss}}\cr
        \noalign{\hrule}%
        \noalign{\vskip\the\lineskip}%
        \mud{\copy\tempa}\cr}}%
                      \tempc=\wd\tempb
                      \advance\tempc by \wd\tempa
                      \divide\tempc by 2 }

\def\inrulean#1#2#3{{\inruleanhelp{#1}{#2}{#3}%
                     \hbox to \wd\tempa{\hss \box\tempb \hss}}}

\def\linrulean#1#2#3{{\inruleanhelp{#1}{#2}{#3}%
                      \hbox to \tempc{\hss \box\tempb}}}

\def\rinrulean#1#2#3{{\inruleanhelp{#1}{#2}{#3}%
                      \hbox to \tempc{\box\tempb \hss}}}

\def\ginrulean#1#2#3{{\inruleanhelp{#1}{#2}{#3}%
                     \hbox to .5\linewidth{\hfil \box\tempb \hfil}}}

\def\ginruleanl#1#2#3{{\inruleanchelp{#1}{#2}{#3}%
                       \hbox to .5\linewidth{\hfil
                         \box\tempb\hskip\wd\tempd \hfil}}} 

\def\inrulem#1#2#3{{\inrulemhelp{#1}{#2}{#3}%
                    \hbox to \wd\tempa{\hss \box\tempb \hss}}}

\def\linrulem#1#2#3{{\inrulemhelp{#1}{#2}{#3}%
                      \hbox to \tempc{\hss \box\tempb}}}

\def\rinrulem#1#2#3{{\inrulemhelp{#1}{#2}{#3}%
                      \hbox to \tempc{\box\tempb \hss}}}

\def\ginrulem#1#2#3{{\inrulemhelp{#1}{#2}{#3}%
                     \hbox to .5\linewidth{\hfil \box\tempb \hfil}}}

\def\ginruleml#1#2#3{{\inrulemchelp{#1}{#2}{#3}%
                     \hbox to .5\linewidth{\hfil
                       \box\tempb \hskip\wd\tempd \hfil}}}

\def\inrulebn#1#2#3#4{\inrulean{#1\qquad\qquad #2}{#3}{#4}}

\def\linrulebn#1#2#3#4{\linrulean{#1\qquad\qquad #2}{#3}{#4}}

\def\rinrulebn#1#2#3#4{\rinrulean{#1\qquad\qquad #2}{#3}{#4}}

\newbox\gappremises

\def\gapbuild#1{\if *#1*\let\next=\relax\else
        \setbox\gappremises=\hbox{\ifvoid\gappremises\else
                                         \unhbox\gappremises \qquad\fi
                                  \mud{#1}}%
        \let\next=\gapbuild\fi\next}

\def\gaphelp#1#2#3{\setbox\tempa=\hbox{$\displaystyle{\mathstrut #2}$}%
                   \setbox\tempb=\vbox{\lineskip=2pt%
 \halign{##\cr
        \mud{#1}\cr
        \noalign{\hrule height 0pt}%
        \mud{#3}\cr
        \noalign{\hrule height 0pt}%
        \noalign{\vskip\the\lineskip}%
        \mud{\copy\tempa}\cr}}%
                      \tempc=\wd\tempb
                      \advance\tempc by \wd\tempa
                      \divide\tempc by 2 }

\def\gapchelp#1#2#3{\setbox\tempa=\hbox{$\displaystyle{\mathstrut #2}$}%
                   \setbox\tempb=\hbox{$\vcenter{\lineskip=2pt%
 \halign{##\cr
        \mud{#1}\cr
        \noalign{\hrule height 0pt}%
        \mud{#3}\cr
        \noalign{\hrule height 0pt}%
        \noalign{\vskip\the\lineskip}%
        \mud{\copy\tempa}\cr}}$}%
                      \tempc=\wd\tempb
                      \advance\tempc by \wd\tempa
                      \divide\tempc by 2 }

\def\inrulegap#1#2#3{{\gaphelp{#1}{#2}{#3}%
                      \hbox to \wd\tempa{\hss \box\tempb \hss}}}

\def\ginrulegap#1#2#3{{\gaphelp{#1}{#2}{#3}%
                       \hbox to 0.5\linewidth{\hfil \box\tempb \hfil}}}

\def\inrulewjanhelp#1#2{\setbox\tempa=\hbox{$\displaystyle{\mathstrut #2}$}%
                        \setbox\tempb=\vbox{\halign{##\cr
        \mud{#1}\cr
        \noalign{\vskip\the\lineskip}%
        \noalign{\vskip 1pt}%
        \noalign{\hrule height 0pt}%
        \noalign{\hrule}%
        \noalign{\vskip\the\lineskip}%
        \mud{\copy\tempa}\cr}}%
                      \tempc=\wd\tempb
                      \advance\tempc by \wd\tempa
                      \divide\tempc by 2 }

\def\inrulewjan#1#2{{\inrulewjanhelp{#1}{#2}%
                     \hbox to \wd\tempa{\hss \box\tempb \hss}}}

\long\def\ignore#1{}

\newenvironment{proof}
     {\begin{trivlist}\item[]{\bf Proof. }}%
     {\\* \hspace*{\fill} $\Box$\end{trivlist}}

     {\\* \hspace*{\fill} $\Box$\end{trivlist}}

\newcounter{exno}

\newcounter{defno}

\newtheorem{lemma}{Lemma}

\newtheorem{theorem}[lemma]{Theorem}

\newcommand{\sep}{\;\vert\;}

\newcommand{\oprove}{\vdash\kern-.6em\lower.5ex\hbox{$\scriptstyle O$}\,}
\newcommand{\cprove}{\vdash\kern-.6em\lower.5ex\hbox{$\scriptstyle C$}\,}
\newcommand{\iprove}{\vdash\kern-.6em\lower.5ex\hbox{$\scriptstyle I$}\,}

\newcommand{\all}{\forall}
\newcommand{\some}{\exists}

\newcommand{\somex}[1]{\some#1\,}
\newcommand\allx[1]{\all#1\,}

\newcommand{\subs}[3]{[#1/#2]#3}

\newcommand{\ie}{{\em i.e.}}

\newcommand{\mthcontr}{\mbox{\rm contr-R}}
\newcommand{\mthcontl}{\mbox{\rm contr-L}}
\newcommand{\contl}{$\mbox{\rm contr-L}$}
\newcommand{\contr}{$\mbox{\rm contr-R}$}
\newcommand{\botr}{$\bot$-R} 
\newcommand{\mthbotr}{\bot\mbox{\rm -R}}
\newcommand{\andl}{$\land$-L} 
\newcommand{\mthandl}{\land\mbox{\rm -L}} 
\newcommand{\mthandlstar}{\land\mbox{\rm -L}^*} 
\newcommand{\andlstar}{$\land$-L$^*$}
\newcommand{\andr}{$\land$-R}
\newcommand{\mthandr}{\land\mbox{\rm -R}} 
\newcommand{\orl}{$\lor$-L}
\newcommand{\mthorl}{\lor\mbox{\rm -L}} 
\newcommand{\orlg}{$\lor$-L$_G$}
\newcommand{\mthorlg}{\lor\mbox{\rm -L}_G} 
\newcommand{\orr}{$\lor$-R}
\newcommand{\orrstar}{$\lor$-R$^*$}
\newcommand{\mthorr}{\lor\mbox{\rm -R}}
\newcommand{\mthorrstar}{\lor\mbox{\rm -R}^*}
\newcommand{\impl}{$\supset$-L}
\newcommand{\implstar}{$\supset$-L$^*$}
\newcommand{\implistar}{$\supset$-L$^*_I$}
\newcommand{\mthimplstar}{\supset\!\mbox{\rm -L}^*}
\newcommand{\mthimplistar}{\supset\!\mbox{\rm -L}^*_I}
\newcommand{\mthimpl}{\supset\!\mbox{\rm -L}}
\newcommand{\impr}{$\supset$-R}
\newcommand{\mthimpr}{\supset\!\mbox{\rm -R}}
 
\newcommand{\alll}{$\forall$-L} 
\newcommand{\alllstar}{$\forall$-L$^*$} 
\newcommand{\mthalll}{\forall\mbox{\rm -L}} 
\newcommand{\mthalllstar}{\forall\mbox{\rm -L}^*} 
\newcommand{\allr}{$\forall$-R}
\newcommand{\mthallr}{\forall\mbox{\rm -R}}
\newcommand{\somel}{$\exists$-L}
\newcommand{\mthsomel}{\exists\mbox{\rm -L}}
\newcommand{\somer}{$\exists$-R}
\newcommand{\somerstar}{$\exists$-R$^*$}
\newcommand{\mthsomer}{\exists\mbox{\rm -R}}
\newcommand{\mthsomerstar}{\exists\mbox{\rm -R}^*}
\newcommand{\resg}{\mbox{\rm res}$_G$}
\newcommand{\mthresg}{\mbox{\rm res}_G} 

\newcommand{\sequent}[2]{\hbox{{$#1\ 
      \longrightarrow\ #2$}}}

\newcommand{\Ibf}{{\bf I}}
\newcommand{\Ibfplus}{{\bf I}$^+$}
\newcommand{\Ibfstar}{{\bf I}$^*$}
\newcommand{\Ibfgprime}{{\bf I$^\prime_G$}}
\newcommand{\Ibfg}{{\bf I}$_G$}
\newcommand{\Obfg}{{\bf O}$_G$}
\newcommand{\Cbf}{{\bf C}}
\newcommand{\Cbfplus}{{\bf C}$^+$}
\newcommand{\Cbfstar}{{\bf C}$^*$}
\newcommand{\Obf}{{\bf O}}

\begin{document}

\begin{center}
{\large\bf Correspondences between Classical, \\
           Intuitionistic and Uniform Provability}\\[7pt]
{\it Gopalan Nadathur}\\[5pt]
Department of Computer Science\\
University of Chicago\\
Ryerson Hall\\
1100 E 58th Street\\
Chicago, IL 60637\\[5pt]
Phone Number: (773)-702-3497\\[5pt]
Fax Number: (773)-702-8487\\[5pt]
Email: \verb+gopalan@cs.uchicago.edu+
\end{center}

\bigskip\bigskip

\begin{abstract}
\noindent Based on an analysis of the inference rules used, we provide
a characterization of the situations in which classical
provability entails intuitionistic provability. We then examine the
relationship of these derivability notions to uniform provability, a
restriction of intuitionistic provability that embodies a special form
of goal-directedness. We determine, first, the circumstances in which
the former relations imply the latter. Using this result, we identify
the richest versions of the so-called abstract logic programming
languages in classical and intuitionistic logic. We then study the
reduction of classical and, derivatively, intuitionistic provability
to uniform provability via the addition to the assumption set of the
negation of the formula to be proved. Our focus here is on
understanding the situations in which this reduction is
achieved. However, our discussions indicate the structure of a proof
procedure based on the reduction, a matter also considered explicitly
elsewhere.
\end{abstract}

\bigskip


\section{Introduction}\label{sec:intro}

We address three questions pertaining to derivability relations over
sequents in this paper. The first of these concerns the correspondence
between classical and intuitionistic provability. It is well known
that the former is a stronger relation than the latter: while every
intuitionistic proof is also a classical one, there are some sequents
that are derivable only in classical logic. However, it is possible in
principle to obtain the reverse correspondence by restricting the
syntax of formulas considered or the kinds of inference rules used in
a classical proof. We examine this possibility here. In particular, we
provide a characterization at the level of inference rule usage of the
situations in which classical provability implies intuitionistic
provability. Our analysis is ``coarse-grained'' in that it pays
attention only to the inference rules used, and not to their
interaction in particular proofs as is done in a restricted setting
in \cite{PRW96cade}, but it is complete at this level of
granularity. While our study is one that has been independently
conducted, results similar to ours have previously been obtained by
Orevkov \cite{Orevkov71} as we discuss in
Section~\ref{sec:classandint}. The results that we present have uses
in proof search. One possible application is that it permits
intuitionistic proof procedures to be employed in settling questions
of classical validity in special situations. This approach has
benefits and has also been employed in the past: for example, it
underlies the procedure commonly used relative to Horn clause logic
with the virtue that proof search at any point is driven by a {\it
single} goal formula. Another application of our observations is that
it supports the use of classical principles in intuitionistic proof
search. Thus, the treatment of quantifier dependencies can, in special
circumstances, be achieved by a static (dual) Skolemization process
instead of a costly dynamic accounting mechanism.

The second question we consider concerns the correspondence between
classical and intuitionistic provability on the one hand and uniform
provability on the other. Uniform proofs as identified in
\cite{MNPS91} are intuitionistic proofs restricted so as to capture a 
goal-directedness in proof search. One reason for
interest in this category of proofs is that it provides a framework
for interpreting the logical symbols in the formulas being proved as
primitives for directing search and the inference rules pertaining to
these symbols as specifications of their search semantics. This
viewpoint has been exploited in \cite{MNPS91} in describing a
proof-theoretic foundation for logic programming. By its very
definition, uniform provability is a less inclusive relation than
either classical or intuitionistic provability. However, by
a suitable restriction of the context, it is possible to obtain a
correspondence between these three relations. We provide, once again,
a complete characterization at the level of inference rule usage of
the situations in which intuitionistic provability entails and uniform
provability. When combined with the earlier result, this analysis yields
a similar characterization relative to classical logic. As one
application of these observations, they enable us to identify the
richest possible logic programming languages within classical and
intuitionistic logic; our remarks relative to intuitionistic logic are
similar to those in \cite{Har94jlc}.

The final question we consider concerns the reduction of classical and
intuitionistic provability to uniform provability. Efficient
procedures can be designed for searching for uniform proofs. Towards
exploiting this possibility, it is worth considering a modification of
the given formula or sequent in a way that does not alter the original
derivability question but, nevertheless, succeeds in reducing it to
one of uniform provability. One such modification that has been
studied in the past is the addition of the negation of the formula
that is to be proved to the assumptions \cite{NL95lics,Nad96jlc}. This
transformation is sound with respect to classical
logic. We characterize the situations in which it also achieves the
desired reduction. Since the transformation can be applied to
intuitionistic provability without loss of soundness whenever this
notion coincides with classical provability, we obtain information
indirectly about the reducibility in this case as well.

\section{Logical preliminaries}\label{sec:basics}

We will work within the framework of a first-order logic in this
paper. The logical symbols that we assume as primitive are $\top$,
$\bot$, $\land$, $\lor$, $\supset$, $\some$, and $\all$. The first two
symbols in this collection denote the tautologous and the contradictory
propositions, respectively. Note that we consider these logical
constants to be distinct from atomic formulas.
Negation is a defined notion in our language, $\neg A$ being an
abbreviation for $(A \supset \bot)$.   

\begin{figure*}[top]

\begin{center}
  \mbox{\inrulean{\sequent{B,B,\Gamma}{\Delta}}
    {\sequent{B,\Gamma}{\Delta}} {\mthcontl}} \qquad\qquad\qquad\qquad
  \mbox{\inrulean{\sequent{\Gamma}{\Delta,B,B}} {\sequent{\Gamma}{\Delta,B}} {\mthcontr}}  \end{center}

\medskip

\begin{center}
    \mbox{\inrulean{\sequent{\Gamma}{\Delta,\bot}}
             {\sequent{\Gamma}{\Delta,D}}
             {\mthbotr}}
\end{center}

\medskip

\begin{center}
    \mbox{\inrulean{\sequent{B, \Gamma}{\Delta}}
             {\sequent{B\land D,\Gamma}{\Delta}}
             {\mthandl}}
    \qquad\qquad\qquad\qquad
    \mbox{\inrulean{\sequent{D, \Gamma}{\Delta}}
             {\sequent{B\land D,\Gamma}{\Delta}}
             {\mthandl}}
\end{center}

\medskip

\begin{center}
   \mbox{
    \inrulebn{\sequent{B,\Gamma}{\Delta}}
             {\sequent{D,\Gamma}{\Delta}}
             {\sequent{B\lor D,\Gamma}{\Delta}}
             {\mthorl}
        }
\end{center}

\medskip

\begin{center}
    \mbox{\inrulebn{\sequent{\Gamma}{\Delta,B}}
             {\sequent{\Gamma}{\Delta,D}}
             {\sequent{\Gamma}{\Delta,B\land D}}
             {\mthandr}}
\end{center}

\begin{center}
    \mbox{\inrulean{\sequent{\Gamma}{\Delta,B}}
             {\sequent{\Gamma}{\Delta,B\lor D}}
             {\mthorr}}
    \qquad\qquad\qquad
    \mbox{\inrulean{\sequent{\Gamma}{\Delta,D}}
             {\sequent{\Gamma}{\Delta,B\lor D}}
             {\mthorr}}
\end{center}

\medskip

\begin{center}
    \mbox{\inrulebn{\sequent{\Gamma}{\Delta,B}}
             {\sequent{D,\Gamma}{\Theta}}
             {\sequent{B\supset D,\Gamma}{\Delta,\Theta}}
             {\mthimpl}}
    \qquad\qquad\qquad\qquad
    \mbox{\inrulean{\sequent{B,\Gamma}{\Delta,D}}
             {\sequent{\Gamma}{\Delta, B\supset D}}
             {\mthimpr}}
\end{center}

\medskip

\begin{center}
    \mbox{\inrulean{\sequent{\subs{t}{x} B,\Gamma}{\Delta}}
             {\sequent{\allx{x} B,\Gamma}{\Delta}}
             {\mthalll}}
    \qquad\qquad\qquad
    \mbox{\inrulean{\sequent{\Gamma}{\Delta, \subs{t}{x}B}}
             {\sequent{\Gamma}{\Delta, \somex{x} B}}
             {\mthsomer}}
\end{center}

\medskip

\begin{center}
    \mbox{\inrulean{\sequent{\subs{c}{x}B,\Gamma}{\Delta}}
             {\sequent{\somex{x}B,\Gamma}{\Delta}}
             {\mthsomel}}
    \qquad\qquad\qquad
    \mbox{\inrulean{\sequent{\Gamma}{\Delta,\subs{c}{x}B}}
             {\sequent{\Gamma}{\Delta,\allx{x} B}}
             {\mthallr}}
\end{center}

\medskip

\caption{Rules for deriving sequents \label{fig:seqrules}}
\end{figure*}

Notions of derivation that are of interest to us are formalized by
sequent calculi. A sequent in our context is a pair of {\it multisets}
of formulas. Assuming that $\Gamma$ and $\Delta$ are
its elements, the pair is written as $\sequent{\Gamma}{\Delta}$ and
$\Gamma$ and $\Delta$ are referred to as its antecedent and succedent,
respectively. Such a sequent is an axiom if either $\top \in \Delta$
or for some $A$ that is either $\bot$ or an atomic
formula, it is the case that $A \in \Gamma$ and $A \in \Delta$. The
rules that may be 
used in constructing sequent proofs are those that can be obtained
from the schemata shown in Figure~\ref{fig:seqrules}. In these
schemata, $\Gamma$, $\Delta$ and $\Theta$ stand for multisets of
formulas, $B$ and $D$ stand for formulas, $c$ stands for a constant,
$x$ stands for a variable and $t$ stands for a term. The notation $B,
\Gamma$ ($\Delta, B$) is used here for a multiset containing the
formula $B$ whose remaining elements form the multiset $\Gamma$
(respectively, $\Delta$). Further, expressions of the form
$\subs{t}{x}B$ are used to denote the result of replacing all free
occurrences of $x$ in $B$ by $t$, with bound variables being renamed
as needed to ensure the logical correctness of these
replacements. There is the usual proviso with respect to the rules
produced from the schemata \somel\ and
\allr: the constant that replaces $c$ should not appear in the
formulas that form the lower sequent. 
A {\it contraction} rule is one that is obtained from either the
\contl\ or the \contr\ schema. 
All other rules are referred to as {\it operational} rules and the
formula in the lower sequent that is explicitly affected by such a
rule is called its {\it principal} formula. Finally, we refer to
\contl\ and the operational rules whose principal formulas are in the
antecedent of the lower sequent as {\it left} rules and to the
remaining rules as {\it right} rules. 

We are interested in three notions of derivability for sequents
of the form $\sequent{\Gamma}{\Delta}$. A \Cbf-proof for such a
sequent is a derivation obtained by making arbitrary uses of the
inference rules. 
\Ibf-proofs are \Cbf-proofs in which every sequent has exactly one
formula in its succedent. Notice that, by this stipulation, $\Delta$
must itself consist of a single formula.  Finally, a 
{\it uniform proof}\ or \Obf-proof is an \Ibf-proof in which any
sequent that has a non-atomic formula distinct from $\bot$ in its
succedent occurs only as the lower sequent of an inference rule that
introduces the top-level logical symbol of that formula. 

In the case that $\Delta$ is a single formula, we shall write $\Gamma
\cprove \Delta$, $\Gamma \iprove \Delta$ and $\Gamma \oprove \Delta$
to indicate the existence of, respectively, a \Cbf-proof, an
\Ibf-proof and an \Obf-proof for $\sequent{\Gamma}{\Delta}$. 
The first two notions correspond to classical and intuitionistic
provability respectively. The sequent calculi that we have used here
to characterize these derivability relations are transparently related
to those in \cite{Prawitz65}: we have treated antecedents and
succedents as multisets rather than sets but have added the
contraction rules to realize arbitrary multiplicity of formulas and,
in the intuitionistic setting, we do not permit sequents of the form 
$\sequent{\Gamma}{}$ that are not derivable in the system of
\cite{Prawitz65}. Uniform provability corresponds to the
existence of an \Obf-proof. This notion indicates the possibility for
a goal-directedness in the search 
for a derivation, with the top-level structure of the formula in the
succedent controlling the next step in the search at each stage. 

We observe certain properties of our derivation calculi for classical
and intuitionistic logic that will be used in later sections. First,
any sequent in which the antecedent and succedent have a common
formula has a \Cbf-proof and, if the succedent has a single element,
an \Ibf-proof. Thus, a modification to our calculi that considers all
such sequents to be axioms does not change the set of provable
sequents. 
The second observation concerns the so-called {\it Cut} inference
rules that are obtained from the schemata

\begin{center}
    \mbox{\inrulewjan{\sequent{\Gamma_1}{\Delta_1,B}\qquad\qquad
             \sequent{B,\Gamma_2}{\Delta_2}}
             {\sequent{\Gamma_1,\Gamma_2}{\Delta_1,\Delta_2}}
             }
\end{center}

\noindent Notice that in generating {\it Cut} rules in the intuitionistic
context, $\Delta_1$ must be instantiated by an empty multiset and
$\Delta_2$ by a singleton multiset. Now, these {\it Cut} rules are
admissible with respect to classical and intuitionistic provability as
formulated here, \ie, the same set of sequents have \Cbf-proofs and
\Ibf-proofs even if we allow these additional rules to be used in
derivations.  This property can be demonstrated by describing a
procedure for eliminating occurrences of the {\it Cut} rules from any
given derivation. An examination of a typical such procedure---for
example, the procedure contained in \cite{Gentzen35}---actually allows
a stronger conclusion to be drawn: a derivation that uses {\it Cut}
rules, contraction rules and operational inference rules obtained from
a restricted subset of the schemata in Figure~\ref{fig:seqrules} can
be transformed into one in which only contraction rules and
operational rules obtainable from the restricted
schemata set appear. Furthermore, this property holds even when the
notion of an axiom is strengthened as described earlier in this
paragraph.

\section{\bf Building contraction into other inference rules}\label{sec:contr}

\noindent The contraction rules allow for a profligate multiplicity of
formulas. The necessary multiplicity can be characterized more
precisely by identifying derived forms of some of the operational
rules that incorporate contraction into their structure and thereby
permit the contraction rules themselves to be omitted from the
calculus. We describe below a convenient form of these
derived rules that is presented, for instance,  in
\cite{Dra79}.

We consider first the case for classical provability. The new rules
that are of interest are those obtained from the following schemata:
\begin{center}
    \mbox{\inrulean{\sequent{A,B,\Gamma}{\Delta}}
             {\sequent{A\land B,\Gamma}{\Delta}}
             {\mthandlstar}}
    \qquad\qquad\qquad
    \mbox{\inrulean{\sequent{\Gamma}{\Delta,A,B}}
               {\sequent{\Gamma}{\Delta,A\lor B}}
             {\mthorrstar}}

\medskip

    \mbox{\inrulebn{\sequent{\Gamma}{B, \Delta}}
             {\sequent{D,\Gamma}{\Delta}}
             {\sequent{B\supset D,\Gamma}{\Delta}}
             {\mthimplstar}}

\medskip

\mbox{\inrulean{\sequent{\allx{x} P,\subs{t}{x} P,\Gamma}{\Delta}}
               {\sequent{\allx{x} P,\Gamma}{\Delta}}
               {\mthalllstar}}
    \qquad\qquad\qquad
\mbox{\inrulean{\sequent{\Gamma}{\Delta,\exists x P,\subs{t}{x} P}}
               {\sequent{\Gamma}{\Delta,\exists x P}}
               {\mthsomerstar}}
\end{center}
These rules are obviously derived ones: a \Cbf-proof of the lower
sequent of each rule can be obtained from \Cbf-proof(s) of the upper
sequent(s) by using an instance of the `asterisk-less' version of the
schema followed by 
some number of contraction rules. By a \Cbfplus-proof let us mean a
derivation constructed in a calculus obtained from the one for
\Cbf-proofs by replacing the rules \andl, \orr, \impl, \alll\ and
\somer\ with the ones obtained from the schemata above. It is then
easily seen that a sequent has a \Cbf-proof if and only if it has a
\Cbfplus-proof. 

Let an \Cbfstar-proof be a \Cbfplus-proof in
which contraction rules are not used. Our objective is to show that a
sequent has a \Cbfstar-proof whenever it has a \Cbfplus-proof, \ie,
contraction can be eliminated from \Cbf-proofs under the described
strengthening of the \andl, \orr, \impl, \alll\ and \somer\
rules. This can be done through the following sequence of steps that
culminate in Theorem~\ref{thm:sanscontraction}. 

\begin{lemma}\label{lem:anyrule}
Let $\sequent{\Gamma}{\Delta}$ have a \Cbfstar-proof of height
$h$ and let all references to rules be to ones for constructing
\Cbfstar-proofs.  

\begin{enumerate}
\item For any single upper sequent rule of the form
\begin{center}
\mbox{\inrulewjan{\sequent{\Gamma'}{\Delta'}}
               {\sequent{\Gamma}{\Delta}}}
\end{center}
it is the case that   $\sequent{\Gamma'}{\Delta'}$ has a
\Cbfstar-proof of height at most $h$.

\item For any rule with two upper sequents of the form
\begin{center}
\mbox{\inrulewjan{\sequent{\Gamma'}{\Delta'}\qquad\qquad
 \sequent{\Gamma''}{\Delta''}}
 {\sequent{\Gamma}{\Delta}}}
\end{center}
it is the case that both $\sequent{\Gamma'}{\Delta'}$
  and $\sequent{\Gamma''}{\Delta''}$ have \Cbfstar-proofs of height at
  most $h$.
\end{enumerate}

\end{lemma}

\begin{proof} The cases for \alllstar\ and \somerstar\ follow from
  observing that if a \Cbfstar-proof exists for a certain sequent,
  then there is a derivation of similar structure and identical height
  for any sequent obtained from it by adding formulas to the
  antecedent or succedent. The remaining cases, that show the
  invertibility of all the other rules, are covered by the
  Inversion Lemma of \cite{TS96}. 
\end{proof}

Let $\Sigma$ be a multiset of formulas. We use the notation
$\hat{\Sigma}$ to denote the {\it set} of formulas appearing in $\Sigma$.

\begin{lemma}\label{lem:superset}
  Let $\sequent{\Gamma}{\Delta}$ have a \Cbfstar-proof of
  height $h$ and let $\Gamma'$ and $\Delta'$ be such that
  $\hat{\Gamma} \subseteq 
  \hat{\Gamma'}$ and $\hat{\Delta} \subseteq \hat{\Delta'}$. Then
  $\sequent{\Gamma'}{\Delta'}$ has a \Cbfstar-proof of height at most
  $h$. 
\end{lemma}

\begin{proof} By an induction on the height of the derivation for
  $\sequent{\Gamma}{\Delta}$. We leave the reader to fill out the
  details, perhaps by consulting \cite{TS96}. 
\end{proof}

\begin{theorem}\label{thm:sanscontraction}
A sequent $\sequent{\Gamma}{\Delta}$ has a \Cbf-proof if and only
if it has a \Cbfstar-proof. 
\end{theorem}

\begin{proof} It suffices to show
  that $\sequent{\Gamma}{\Delta}$ has a \Cbfplus-proof if and only if
  it has a \Cbfstar-proof. The `if' direction is obvious. For the
  other direction we use an induction on the number of contractions in
  the \Cbfplus-proof. If there are none, then we already have a
  \Cbfstar-proof. Otherwise, we find a contraction   that is the first
  one in the derivation in the path from an axiom to the final
  sequent. By Lemma~\ref{lem:superset}, this contraction can be
  dispensed with, yielding a \Cbfplus-proof with one less contraction. 
\end{proof}

Theorem~\ref{thm:sanscontraction} and Lemma~\ref{lem:anyrule}  provide
the basis for a proof procedure for classical logic that is worth
noting. All the rules that may be used in \Cbfstar-proofs that are
distinct from the \alll\ and \somer\ rules have an upper sequent or two upper
sequents with fewer logical symbols 
than those in the lower sequent. We may therefore use these rules
repeatedly in a terminating process to reduce a given sequent to a new
set of 
sequents for which derivations must be constructed. If every sequent
in the set so produced is an axiom, then we will have established
classical provability. If at least one of the new sequents is not
an axiom and cannot be the lower sequent of either a \alllstar\ or a
\somerstar, then the original sequent can have no
\Cbf-proof. Otherwise each non-axiom sequent is reduced by a
simultaneous use of all the \alllstar\ and \somerstar\ rules that are
applicable to it and the process is repeated. 
The procedure as presently stated is not quite practical since the use of
a \alllstar\ or a \somerstar\ rule also involves picking the `right'
instantiation term. However, this choice can be delayed by introducing
instead a variable that may be instantiated later and determining
bindings for such 
variables by using unification when checking if a sequent is an
axiom. Unification must, of course, not lead to the constraint on
constants in the \somel\ and \allr\ rules to be violated. The best way
to achieve this effect in the classical setting is to transform the
original sequent by a process referred to as Herbrandization in
\cite{Shankar92} that eliminates at the outset all quantifiers that
might require a \somel\ and \allr\ to be used in proof search. 

Contraction is applicable only to antecedent formulas in the 
intuitionistic setting. The essential uses of these rules occur in
conjunction with the \andl, \alll, and \impl\ rules. To realize the
effects of these uses, we may replace the \andl\ and the
\alll\ schemata by \andlstar\ and \alllstar\ respectively, and \impl\ by
the following: 
\begin{center}
    \mbox{\inrulebn{\sequent{B \supset D,\Gamma}{B}}
             {\sequent{D,\Gamma}{\Delta}}
             {\sequent{B\supset D,\Gamma}{\Delta}}
             {\mthimplistar}}
\end{center}
Notice that, in contrast to the situation in classical logic, the
present modification to the \impl\ rules incorporates a contraction in
the {\it antecedent}. 

We refer to a derivation constructed in the calculus for
\Ibf-proofs with the indicated replacements for the \andl, \alll\ and
\impl\ rules as an \Ibfplus-proof. If \contl\ is not used in such a
derivation, we shall call it an \Ibfstar-proof. The adequacy
of \Ibfstar-proofs in settling questions of intuitionistic provability
is stated in the following theorem whose proof may be modelled on the
arguments in \cite{TS96}. 

\begin{theorem}\label{thm:sanscontint}
A sequent $\sequent{\Gamma}{F}$ has a \Ibf-proof if and only
if it has a \Ibfstar-proof. 
\end{theorem}

Once again, Theorem~\ref{thm:sanscontint} has content that can be
utilized in structuring proof
search in intuitionistic logic. However, there are important
differences from the classical case. First, the static Herbrandization
step is not sound in the new setting \cite{Nad92int, Shankar92}. 
An alternative approach that can be used in this case is to treat the
\somel\ and \allr\ rules explicitly in proof search and to employ a
dynamic form of Herbrandization to ensure that the required
constraints are satisfied by quantifier instantiation terms
\cite{Fitting90,Nad92int,Shankar92}. 
Second, a detailed analysis reveals that the process for reducing
sequents must delay consideration of the rules \implistar, \orr\ and
\somer\ in addition to \alllstar.\footnote{There is actually another
  problem with regard to the \implistar\ rules: the principal formula
  of this rule appears again in the left upper sequent and so it is
  not certain that a use of the rule will produce less `complex'
  sequents.  
Dyckhoff \cite{dyckhoff92} and Hudelmaier \cite{Hudelmaier90} 
have proposed alternative sequent calculi for propositional logic that
overcomes this problem but, to our knowledge, no similar calculus has
been described for the situation where quantifiers are included.}
Further, the order in which these rules are eventually considered may
be important and it may be necessary to backtrack over particular
orders of reduction. 

 \section{Correspondence between classical and intuitionistic
provability}\label{sec:classandint} 

It is clear from the definitions that, if $\Gamma$ is a (multi)set of
formulas and $\Delta$ is a single formula, then $\Gamma
\iprove \Delta$ implies $\Gamma \cprove \Delta$. 
The converse is not always true. A `canonical' demonstration
of this fact is obtained by taking $\Gamma$ to be the empty (multi)set
and letting $\Delta$ consist of the formula $((q \supset s) \supset q)
\supset q$. 
However, the truth of the converse and, hence, the equivalence of
classical and intuitionistic provability, can be assured when the
syntax of the assumption and conclusion formulas is restricted in
certain ways. 
We describe these syntactic restrictions in this section. 
Our characterization is based, first of all, on the inference
rules used in a \Cbf-proof and is a complete one at this level: we
identify four classes of \Cbf-proofs determined by the non-use of
certain inference rules and show that (a)~an \Ibf-proof exists for the
final sequent of a \Cbf-proof belonging to any of these classes and
(b)~for each possible way for violating all the restrictions on inference
rule usage, there is a \Cbf-proof with a corresponding final sequent
for which no \Ibf-proof exists.  
Now, the syntactic structure of the formulas in a given sequent
determines the inference rules that can appear in a (cut-free)
\Cbf-proof of that sequent. 
This observation enables us to translate the restriction on inference
rules into the desired syntactic constraints on formulas. 

The following theorem identifies one of the classes of \Cbf-proofs that
are of interest. 

\begin{theorem}\label{thm:improrl}
Let $\sequent{\Gamma}{\Delta}$ have a \Cbf-proof in which no
\impr\ or \orl\ rule is used. Then, for some $G$ in $\Delta$, it is
the case that $\sequent{\Gamma}{G}$ has an \Ibf-proof. In particular,
if $\Delta$ consists of a single formula, then
$\sequent{\Gamma}{\Delta}$ itself has an \Ibf-proof.
\end{theorem}

\begin{proof} We use an induction on the heights of \Cbf-proofs. If
  $\sequent{\Gamma}{\Delta}$ is an axiom, then, clearly, there is a
  $G$ in $\Delta$ such that $\sequent{\Gamma}{G}$ is also an axiom. Thus, the
  theorem is true for \Cbf-proofs of height $1$. If the height of the
  derivation is greater than $1$, we consider each possibility for the
  last rule used. If this is a rule with a single upper sequent, then, by
  assumption, it must be distinct from an \impr. In all the remaining
  cases, the   induction hypothesis combined possibly with a rule
  obtained from the same schema yields the desired conclusion. If the
  last rule has two upper sequents, then, since it is distinct from an
  \orl, it must be either an \impl\ or an \andr. Suppose it is the
  first. Then the derivation at the end has the structure 
\begin{center}
\mbox{\inrulewjan
          {\sequent{\Gamma'}{\Delta_1,B} \qquad\qquad
            \sequent{D,\Gamma'}{\Delta_2}} 
          {\sequent{B \supset D, \Gamma'}{\Delta_1,\Delta_2}}}
\end{center}
where $\Gamma$ is $B\supset D,\Gamma'$ and $\Delta_1$ and $\Delta_2$
constitute a (multiset) partition of $\Delta$. By hypothesis, either 
$\sequent{\Gamma'}{G}$ has an \Ibf-proof for some $G$ in $\Delta_1$ or
$\sequent{\Gamma'}{B}$ has an \Ibf-proof. In the first case, it is
easily seen that $\sequent{B \supset D,\Gamma'}{G}$ also has an
\Ibf-proof. In the second case, we use the hypothesis again to observe
that for some $G$ in $\Delta_2$ it is the case that
$\sequent{D,\Gamma'}{G}$ has an \Ibf-proof. These observations
used together with an \impl\ rule yields the theorem in this case. A
similar argument can be provided when the last rule is an \andr.
\end{proof}

We translate the restriction on proof rules in Theorem~\ref{thm:improrl} 
into restrictions on the syntax of formulas. Consider the classes of
formulas defined by the following mutually recursive syntax rules,
assuming $A$ represents atomic formulas: 
\[ \begin{array}{rcl}
 G & ::= & \top \sep \bot \sep A \sep G\land G \sep G\lor G \sep \allx{x}G
 \sep \somex{x} G \\
 D & ::= & \top \sep \bot \sep A \sep G\supset D \sep D \land D \sep
 \somex{x} D \sep \allx{x} D \end{array} \] 
A sequent in which the succedent consists of a $G$-formula and the
antecedent contains only $D$-formulas is classically provable just in
 case it is intuitionistically provable. We observe that the $G$- and
 $D$-formulas defined here subsume the so-called goal formulas and
 program clauses of Horn clause logic \cite{MNPS91}.

There is an auxiliary utility to Theorem~\ref{thm:improrl}: it has
content relevant to defining a multi-formula succedent sequent
calculus for intuitionistic logic. 
Such a calculus is of interest because it permits a postponement in
proof search of decisions about which disjunct of a disjunctive 
formula in the succedent is to be chosen.
Consider the calculus for constructing \Cbfstar-proofs
with the \orl\ and \impr\ rules replaced, respectively, with ones
obtained from the following schemata: 
\begin{center}
   \mbox{
    \inrulewjan{\sequent{B,\Gamma}{F}\qquad\qquad\sequent{D,\Gamma}{F}}
             {\sequent{B\lor D,\Gamma}{\Delta,F}}
        }

\medskip

    \mbox{\inrulewjan{\sequent{B,\Gamma}{D}}
             {\sequent{\Gamma}{\Delta, B\supset D}}
            }
\end{center}
A sequent of the form $\sequent{\Gamma}{\Delta}$ in which $\Delta$ is
a singleton multiset has a derivation in this calculus if and only if it
has an \Ibf-proof; the `if' direction is obvious and the `only if'
direction follows from Theorem~\ref{thm:improrl} and an easy induction
on the number of occurrences of the `new' \orl\ and \impr\ rules in
the given \Cbfstar-proof. We may also allow $\Delta$ to contain more
than one formula by interpreting it as the disjunction of these
formulas. We note, however, that the modification to the \orl\ schema is
essential even under such an interpretation for quantificational logic: 
without this modification, the sequent $\sequent{\allx{x}(p(x) \lor
  q)}{(\allx{x}p(x)) \lor q}$ would, for example, have a derivation
even though it has no \Ibf-proof. 

The following theorem identifies a second interesting class of
\Cbf-proofs. 

\begin{theorem}\label{thm:imprallr}
Let $\Gamma$ be a multiset of formulas and let $B_1,\ldots,B_n$ be
formulas such that $\sequent{\Gamma}{B_1,\ldots,B_n}$ has a \Cbf-proof
in which no \impr\ or \allr\ rule is used. Then the sequent
$\sequent{\Gamma}{B_1 \lor \ldots \lor B_n}$ has an \Ibf-proof. In the
case that $n= 1$, $\sequent{\Gamma}{B_1}$ has an \Ibf-proof. 
\end{theorem}

\begin{proof}
\noindent This theorem can be proved, once again, by an induction on
the heights of \Cbf-proofs. We do not provide an explicit proof here,
noting only that an argument that is similar to, but simpler than,
that for Theorem~\ref{thm:impralll} below suffices. In particular, a
complication arises in the (inductive) proof of
Theorem~\ref{thm:impralll} from having to consider a \allr\ as the
last rule in the \Cbf-proof of
$\sequent{\Gamma}{B_1,\ldots,B_n}$. The premise of the present theorem
rules out this possibility. There is an additional case that has to be
considered here in that a \alll\ rule could be the last one
used. However, the argument for this case is a relatively simple one. 
\end{proof}

Following earlier lines, we can rephrase Theorem~\ref{thm:imprallr}
in terms of a restriction on the syntax of formulas. Consider the
following classes of formulas, assuming, again, that $A$ represents
atomic formulas: 
\[ \begin{array}{rcl}
 G & ::= & \top \sep \bot \sep A \sep G\land G \sep G\lor G \sep \somex{x} G\\
 D & ::= & \top \sep \bot \sep A \sep G\supset D \sep D \land D \sep D
 \lor D \sep \somex{x} D \sep \allx{x} D. \end{array} \]
If $\Gamma$ is a (multi)set of $D$-formulas and $F$ is a $G$-formula,
 then $\Gamma \cprove F$ only if $\Gamma \iprove F$. The classes of
 $G$- and $D$-formulas described by the present rules constitute a
 generalization of similarly named classes in \cite{NL95lics} and have
 been studied there as the basis for disjunctive logic programming
 \cite{LMR92book}. 

Analogously to Theorem~\ref{thm:improrl},
Theorem~\ref{thm:imprallr} can be used to justify a multi-formula
succedent sequent calculus for intuitionistic logic. 
Consider the calculus for constructing \Cbfstar-proofs
with the \allr\ and \impr\ rules replaced, respectively, with ones
obtained from the following schemata: 
\begin{center}
   \mbox{
    \inrulewjan{\sequent{\Gamma}{\subs{c}{x}B}}
             {\sequent{\Gamma}{\Delta,\allx{x} B}}}

\medskip

    \mbox{\inrulewjan{\sequent{B,\Gamma}{D}}
             {\sequent{\Gamma}{\Delta, B\supset D}}
            }
\end{center}
A sequent of the form $\sequent{\Gamma}{\Delta}$ in which $\Delta$ is
a singleton multiset has a derivation in this calculus if and only if it
has an \Ibf-proof. As before, we may also allow $\Delta$ to contain
more than one formula by interpreting it as the disjunction of these
formulas. We note that the calculus that is so described for
intuitionistic logic differs superficially---in particular, only in
the manner in which the logical constant $\bot$ is treated---from the
GHPC calculus of Dragalin \cite{Dra79}.

A third category of \Cbf-proofs is identified in the following
theorem. 

\begin{theorem}\label{thm:impralll}
Let $\Gamma$ be a multiset of formulas and let $B_1,\ldots,B_n$ be
formulas such that $\sequent{\Gamma}{B_1,\ldots,B_n}$ has a \Cbf-proof
in which no \impr\ or \alll\ rule is used. Then the sequent
$\sequent{\Gamma}{B_1 \lor \ldots \lor B_n}$ has an \Ibf-proof. In the
case that $n= 1$, $\sequent{\Gamma}{B_1}$ has an \Ibf-proof. 
\end{theorem}

\begin{proof} It is convenient to prove the theorem assuming
  derivation calculi with the stronger notion of axioms described in
  Section~\ref{sec:basics}, \ie, ones in which any sequent whose
  antecedent and succedent have a common formula is considered an
  axiom. Further, we show a stronger property than that required: If
  $\sequent{\Gamma}{B_1,\ldots,B_n}$ has a \Cbf-proof in which no
  \impr\ or \alll\ rule is used, then
  $\sequent{\Gamma}{B_1\lor\ldots\lor B_n}$ has an \Ibf-proof in which
  no \alll\ rule is used. We prove this property by means of an
  induction on the height of the \Cbf-proof for
  $\sequent{\Gamma}{B_1,\ldots,B_n}$. 

The base case corresponds to $\sequent{\Gamma}{B_1,\ldots,B_n}$ being
  an axiom. In this case, we can construct an \Ibf-proof of
  $\sequent{\Gamma}{B_1 \lor \ldots \lor B_n}$ by using a sequence of
  \orr\ rules below a suitably chosen axiom. Note that no \alll\ rule
  appears in this derivation.  

For the inductive step, we consider the various possibilities for the
last rule used in the derivation. The argument is straightforward for
all permitted left rules, \ie, ones that are distinct from \alll, in
which the 
succedent is identical in the upper and lower sequents---we invoke the
induction hypothesis and use an instance of the same rule schema to
get an \Ibf-proof for $\sequent{\Gamma}{B_1 \lor \ldots \lor B_n}$,
and we note that this derivation must not contain a \alll\ rule
occurrence.  

The only remaining possibility for a left rule is that it is an
\impl. In this case, the derivation at the end has the 
structure
\begin{center}
\mbox{\inrulewjan{\sequent{\Gamma'}{\Delta_1,F}\qquad\qquad
                  \sequent{D,\Gamma'}{\Delta_2}}
                 {\sequent{F \supset D,\Gamma'}{\Delta_1,\Delta_2}}}
\end{center}
where $\Gamma$ is $F\supset D,\Gamma'$ and $\Delta_1$ and $\Delta_2$
constitute some partition of $B_1,\ldots,B_n$. The argument follows
the pattern of that for the 
other left rules in the case that $\Delta_1$ is empty. We therefore
assume that it is nonempty. We also assume that the final sequent
in the derivation has exactly two formulas in the succedent and that
$\Delta_1$ is $B_1$ and $\Delta_2$ is $B_2$; these assumptions are not
critical, and may be dispensed with in a more detailed argument.

Now, using the induction hypothesis, we see that $\sequent{\Gamma'}{B_1
  \lor F}$ and $\sequent{D,\Gamma'}{B_2}$ have \Ibf-proofs in which no
  \alll\ rule is used. From this it follows that 
  $\sequent{F \supset D, \Gamma'}{B_1 \lor F}$ and
  $\sequent{F,D,\Gamma'}{B_2}$ also have such \Ibf-proofs. Using the
  latter, we can construct an \Ibf-proof for $\sequent{B_1 \lor F, F
  \supset D,\Gamma'}{B_1 \lor B_2}$ as follows: 
\begin{center}
\mbox{\inrulebn{\linrulean{\sequent{B_1, F \supset D, \Gamma'}{B_1}}
                          {\sequent{B_1,F \supset D, \Gamma'}{B_1 \lor B_2}}
                          {\mthorr}
               }
               {\rinrulebn{\sequent{F,\Gamma'}{F}}
                          {\rinrulean{\sequent{F,D,\Gamma'}{B_2}}
                                     {\sequent{F,D,\Gamma'}{B_1 \lor B_2}}
                                     {\mthorr}
                          }
                          {\sequent{F,F\supset D,\Gamma'}{B_1 \lor B_2}}
                          {\mthimpl}
               }
               {\sequent{B_1 \lor F, F \supset D,\Gamma'}{B_1 \lor B_2}}
               {\mthorl}
\hspace{2cm}}
\end{center}
Notice that no \alll\ rule appears in this derivation. We
can combine this derivation with the one for $\sequent{F \supset D,
  \Gamma'}{B_1 \lor F}$ by means of a {\it Cut} rule and some
\contl\ rules to get a derivation for $\sequent{F
  \supset D, \Gamma'}{B_1 \lor B_2}$. Finally, by the observation in
Section~\ref{sec:basics}, the {\it Cut} rule can be eliminated from
this derivation to obtain an \Ibf-proof for $\sequent{F \supset D,
  \Gamma'}{B_1 \lor B_2}$ in which no \alll\ rules appear.

To complete the proof, we have to consider the possibility that the
last rule in the \Cbf-proof is a right rule. Suppose that it is, in
fact, a \somer. Then 
the derivation at the end has the following form:
\begin{center}
\mbox{\inrulewjan{\sequent{\Gamma}{B_1, \ldots, B_{i-1}, \subs{t}{x}B'_i,
\ldots, B_n}}
            {\sequent{\Gamma}{B_1, \ldots, B_{i-1}, \somex{x}B'_i, \ldots, B_n}}}.
\end{center}
We have assumed here that $B_i$ is actually a formula of the form
$\somex{x}B'_i$. 
By the induction hypothesis, $\sequent{\Gamma}{B_1 \lor \ldots \lor
  B_{i-1} \lor \subs{t}{x}B'_i \lor \ldots \lor B_n}$ has an
\Ibf-proof. Now, it is easily seen that 
\begin{tabbing}
\quad\=\kill
\> $\sequent{B_1 \lor \ldots \lor B_{i-1} \lor \subs{t}{x}B_i \lor
  \ldots \lor B_n}{B_1 \lor \ldots \lor B_{i-1} \lor \somex{x} B'_i
  \lor \ldots \lor B_n}$
\end{tabbing}
\noindent has an \Ibf-proof in which no \alll\ rule is used. 
The desired conclusion follows in this case first from using a {\it
  Cut} rule and then noting that this rule can be eliminated from the
derivation without introducing any occurrences of the \alll\ rule.  

An argument similar to that for \somer\ can be provided for all other
permitted right rules except \allr. For the case of \allr, we need a
further observation: If a sequent of the form $\sequent{\Gamma}{P_1
  \lor \ldots   \lor \subs{c}{x}P'_i \lor \ldots \lor P_n}$ has an
\Ibf-proof in which no \alll\ rule is used and if $\Gamma$ and
$P_1, \ldots, P'_i,\ldots P_n$ are such that the constant $c$ does not
appear in them, then $\sequent{\Gamma}{P_1 \lor \ldots \lor
  \allx{x}P'_i   \lor \ldots \lor P_n}$  has an \Ibf-proof in which no
\alll\ rule is used. This observation can be established by a routine
induction on the height of the given \Ibf-proof. Further, it can be
used together with the present induction hypothesis to yield an
argument for the only remaining case in the proof of the main claim.
\end{proof}

We can, as usual, rephrase Theorem~\ref{thm:impralll} in terms of a
restriction on the syntax of formulas. Once again, consider the
following classes of formulas, assuming that $A$ represents atomic
formulas: 
\[ \begin{array}{rcl}
 G & ::= & \top \sep \bot \sep A \sep G\land G \sep G\lor G \sep
 \somex{x} G \sep \allx{x} G\\
 D & ::= & \top \sep \bot \sep A \sep G\supset D \sep D \land D \sep D
 \lor D \sep \somex{x} D. \end{array} \]
If $\Gamma$ is a (multi)set of $D$-formulas and $F$ is a $G$-formula,
 then $\Gamma \cprove F$ only if $\Gamma \iprove F$. 

The following theorem identifies a fourth, and final, class of
\Cbf-proofs that are of interest from the perspective of this
section. 

\begin{theorem}\label{thm:implsomerorr}
Let $\Gamma$ be a multiset of formulas and let $B$ be a
formula such that $\sequent{\Gamma}{B}$ has a \Cbf-proof
in which no \impl, \orr\ or \somer\ rule is used. Then 
$\sequent{\Gamma}{B}$ has an \Ibf-proof. 
\end{theorem}

\begin{proof}
We claim that if $\sequent{\Gamma}{B}$ has a \Cbf-proof in which no
\impl, \orr\ or \somer\ rule is used, then this sequent also
has a \Cbfstar-proof in which no \implstar, \orrstar\ or \somerstar\
rule is used. Towards seeing this, we first make the easy
observation that, under the given assumption,
$\sequent{\Gamma}{B}$ must have a \Cbfplus-proof in which the latter
rules do not appear. Now, it is easily determined that the 
\Cbfstar-proofs mentioned in the Lemmas~\ref{lem:anyrule} and
\ref{lem:superset} may be qualified to be ones
in which the \implstar, \orrstar\ and \somerstar\ rules do not
appear. Finally, an argument similar to that provided for
Theorem~\ref{thm:sanscontraction} allows us to conclude that
$\sequent{\Gamma}{B}$ has a \Cbfstar-proof that does not contain any 
\implstar, \orrstar\ or \somerstar\ rules. 

The claim easily yields the theorem: Every sequent in the
\Cbfstar-proof of restricted form must have exactly one formula in the
succedent. Each occurrence of an \andlstar\ and \alllstar\ rule in this
derivation can be eliminated in favor of a \contl\ rule paired with
some number of \andl\ and \alll\ rules, respectively, to produce a
\Cbf-proof. The number of formulas in the succedent of each sequent
remains unchanged by this transformation and so the \Cbf-proof that is
produced is also an \Ibf-proof.
\end{proof}

Towards rephrasing Theorem~\ref{thm:implsomerorr} in terms of a
restriction on formulas, we define the following classes
of formulas, assuming, as usual, that $A$ represents atomic formulas:
\[ \begin{array}{rcl}
 G & ::= & \top \sep \bot \sep A \sep G\land G \sep D\supset G \sep
 \allx{x} G\\
 D & ::= & \top \sep \bot \sep A \sep D \land D \sep D \lor D \sep
 \somex{x} D \sep \allx{x} D. \end{array} \]
It follows from the theorem that if $\Gamma$ is a (multi)set of
 $D$-formulas and $F$  is a $G$-formula, then $\Gamma \cprove F$ only
 if $\Gamma \iprove F$.  

\begin{theorem}\label{thm:ctoicomp}
Theorems~\ref{thm:improrl}-\ref{thm:implsomerorr} provide a
characterization at the level of proof rules of the conditions under
which classical provability implies intuitionistic provability that is
complete in the following sense: for each way of violating all the
restrictions on inference rule usage described in the mentioned
theorems, there is a sequent with a singleton succedent that has a
violating \Cbf-proof but no \Ibf-proof.
\end{theorem}

\begin{proof}
\Cbf-proofs may be categorized into those that do and those that do
not contain occurrences of the \impr\ rules. 

We consider first the collection of \Cbf-proofs in which the \impr\ 
rules are {\it not} used. To violate the restrictions on
proof rule usage contained in
Theorems~\ref{thm:improrl}-\ref{thm:implsomerorr}, a derivation of
this kind must contain at least one occurrence of an \orl, a \allr\
and a \alll\ rule and of either an \impl, an \orr\ or a \somer\
rule. We list sequents below that have \Cbf-proofs satisfying each of
these requirements and note that none of these has an \Ibf-proof:
\begin{tabbing}
\qquad\=\kill
\>$\sequent{( \allx{x}p(x)) \supset q, \allx{x}(p(x) \lor q)}{q}$\\
\>$\sequent{\allx{x}(p(x) \lor q)}{(\allx{x}p(x)) \lor q}$\\
\>$\sequent{\allx{x}\allx{y}(r(x,a) \lor r(y,b))}{\somex{y}\allx{x}r(x,y)}$.
\end{tabbing}
\noindent We assume in these sequents that $q$ is a proposition
symbol, $p$ is a unary predicate symbol, $r$ is a binary predicate
symbol and $a$ and $b$ are constants. 

A \Cbf-proof in which an \impr\ rule is used must also contain an
occurrence of one of the \impl, \orr\ and \somer\ rules in order to
violate the restrictions described in
Theorems~\ref{thm:improrl}-\ref{thm:implsomerorr}. The following
sequents have \Cbf-proofs satisfying each of these requirements:
\begin{tabbing}
\qquad\=\kill
\>$\sequent{(q \supset s) \supset q}{q}$\\
\>$\sequent{}{q \lor (q \supset s)}$\\
\>$\sequent{}{\somex{x}(p(x) \supset p(f(x)))}$
\end{tabbing}
In these sequents, we assume additionally that $s$ is a proposition
symbol and that $f$ is a unary function symbol. It is easily seen that
none of these sequents has an \Ibf-proof, thus verifying the theorem
even in this case. 
\end{proof}

We stress, once again, the observation made in Section~\ref{sec:intro}
that our analysis of the correspondence between classical and
intuitionistic provability is coarse-grained in that it pays attention
only to the rules used in a derivation and not to the particular
interactions between rules in it. Thus, there are sequents
whose only \Cbf-proofs violate all the conditions in
Theorems~\ref{thm:improrl}-\ref{thm:implsomerorr} but which,
nevertheless, have \Ibf-proofs. For example, consider the sequent 
\begin{tabbing}
\qquad\=\kill
\>$\sequent{}{((\allx{x}(r(x,a) \lor r(x,b))) \supset
  ((((\allx{x}\somex{y}r(x,y))\supset q) \supset q) \lor s))}$
\end{tabbing}
\noindent in which we have used the non-logical vocabulary described
in the proof of Theorem~\ref{thm:ctoicomp}. Any \Cbf-proof of this
sequent must use an \orl, an \impr, a \allr, a \alll, an \orr, a
\somer\ and an \impl\ rule. However, this sequent has an \Ibf-proof. 
We note that it is possible to conduct an alternative analysis of
the correspondence between classical and intuitionistic provability
that focuses specifically on the {\it interactions} between the rules
that appear in a derivation. Such a study has, for instance, been
carried out in \cite{PRW96cade} for a propositional logic that has
$\supset$, $\land$ and $\neg$ as its only logical symbols.
An analysis of this sort indicates when a given classical derivation
may be interpreted as having intuitionistic force and may be used in
driving a search for a \Cbf-proof with such a force given one without
it. The results of this section are relevant to such a study in that
they provide insight into the rules between which interactions should
be considered carefully. 

After the completion of this paper, it has come to our attention that a
study similar to the one presented in this section has previously been
conducted by Orevkov \cite{Orevkov71}. In this work, the notion
of a $\sigma$-class is identified as a list of logical symbols with
positive or negative markings. A sequent is said to belong to a given
$\sigma$-class if a logical symbol occurs positively (negatively) in
the sequent 
only if it does not occur with a corresponding positive (negative)
marking in the listing denoting the $\sigma$-class. Viewed
differently, a $\sigma$-class describes a restriction to the
syntax of formulas that are permitted to appear in sequents. A {\it
  completely Glivenko class} is now defined to be a $\sigma$-class
such that any sequent with a singleton succedent belonging to that
class is derivable in classical predicate logic only if it is also
derivable in intuitionistic predicate logic.\footnote{In reality, it
  is predicate logic with {\it equality} that is considered in
  \cite{Orevkov71}, but this appears not to be significant to the
  analysis.} Analogous to our Theorem~\ref{thm:ctoicomp}, Orevkov
provides a complete description of all completely Glivenko
classes. The two characterizations are not exactly identical because
negation is treated in \cite{Orevkov71} as a primitive
symbol. However, a comparison of the results can still be
made. Ignoring the negation symbol, the two characterizations 
coincide. Treating the negation symbol explicitly allows for
distinctions in \cite{Orevkov71} that, in our context, would translate
not into restrictions in rule usage, but into distinguishing different
roles for implication and paying attention to the polarity of
occurrences of $\bot$. 

\section{Relationship to uniform provability}\label{sec:unif}

We consider now the relationship between classical and
intuitionistic provability on the one hand and uniform provability on
the other. Our analysis covers two kinds of questions. First, we
examine restrictions in the syntax of formulas that ensure a
coextensiveness between these different proof relations. 
Following this, we consider the reduction of classical provability
to uniform provability in situations where these relations are {\it
  not} coextensive. 

\subsection{Correspondence with uniform provability}
Our first goal is to describe the sequents for which the
existence of an \Ibf-proof implies the existence of an
\Obf-proof. Since an \Obf-proof is a special case of an \Ibf-proof,
we can combine this characterization with the results of the previous
section to obtain a similar relationship between classical and uniform
provability. 
The following theorem provides the desired characterization in terms of
the inference rules used in the \Ibf-proof. 

\begin{theorem}\label{thm:inttounif}
Let $\Gamma$ be a multiset of formulas and let $G$ be a formula. If
the sequent $\sequent{\Gamma}{G}$ has an \Ibf-proof in which 
\begin{enumerate}
\item either no \orl\ rule is used or no \orr\ and no \somer\ rules
  are used, and 
\item either no \somel\ rule or no \somer\ rule is used,
\end{enumerate}
then it also has a uniform proof. Moreover, this characterization is
tight in that, for each possible way of violating these restrictions,
there is a sequent with an \Ibf-proof but no uniform proof. 
\end{theorem}

\begin{proof}
The first part of the theorem is an immediate consequence of the
permutability properties of inference rules in intuitionistic sequent
calculi established, for instance, in \cite{Kleene52a}. 
To complete the proof of the theorem, we list a suitable set of
sequents:
\begin{tabbing}
\qquad\=\kill
\>$\sequent{p(a) \lor p(b)}{\somex{x}p(x)}$,\\
\>$\sequent{q \lor s}{s \lor q}$, and\\
\>$\sequent{\somex{x}(p(x) \land q)}{\somex{x}p(x)}$.
\end{tabbing}
\noindent We assume here that $q$ and $s$ are proposition symbols, $p$
is a predicate symbol and $a$ and $b$ are constants. None of these
sequents has a uniform proof. However all of them have \Ibf-proofs:
the first has one in which an \orl\ and a \somer\ rule are used, the
second has one in which an \orl\ and an \orr\ rule are used and the
last has one in which a \somel\ and a \somer\ rule are used. 
\end{proof}

The notion of uniform provability is useful in identifying logical
languages that provide a basis for 
programming \cite{MNPS91}. In particular, 
letting $\cal D$ and $\cal G$ denote collections of formulas and
$\vdash$ denote a chosen proof relation, an {\it abstract logic
  programming language} is defined to be a triple $\langle {\cal D},
{\cal G}, \vdash \rangle$ such that, for all finite subsets $\cal P$
of $\cal D$ and all $G \in {\cal G}$, ${\cal P} \vdash G$ if and only
if ${\cal P} \oprove G$. 
In the programming interpretation of such a triple, elements of ${\cal
  D}$ function as program clauses and elements of ${\cal G}$ serve as
queries or goals. 
The virtue of this definition is that it supports a
broad interpretation of logic programming based on a duality in the
meaning of logical symbols: on the one hand, these symbols have a
declarative reading given by the proof relation $\vdash$ and, on the
other, they are accorded a search-related interpretation given by the
rules for introducing each of them on the right in sequent proofs. 

An interesting question is that of how rich the syntax of program
clauses and goals can be in the cases where $\vdash$ is interpreted as
classical or intuitionistic provability. Before answering this
question, we note that these formulas must contain certain syntactic
components in order to be useful for programming: the procedural
interpretation of 
program clauses relies on universal quantification and implications
being permitted at the top-level in these formulas and outermost
existential quantification is important in goals in making
sense of the result of finding a derivation. In light of
Theorem~\ref{thm:inttounif}, the second requirement precludes 
outermost occurrences of disjunction and existential quantification in
program clauses. Thus, if $\vdash$ is interpreted as intuitionistic
provability, the collection of $G$- and $D$-formulas given by the
following syntax rules represent the largest possible classes for
goals and program clauses:
\[ \begin{array}{rcl}
 G & ::= & \top \sep \bot \sep A \sep G\land G \sep G\lor G \sep D
 \supset G \sep \allx{x}G \sep \somex{x} G \\
 D & ::= & \top \sep \bot \sep A \sep G\supset D \sep D \land D \sep
 \sep \allx{x} D \end{array} \] 
We assume, as before, that $A$ represents atomic formulas in these rules.
The only essential difference between the abstract logic programming
 language given by these classes of formulas and intuitionistic
 provability and the language of hereditary Harrop formulas studied in
 \cite{MNPS91} is that the logical constant $\bot$ is permitted to
 appear here in goals and program clauses. 

In the case that classical provability is used instead to clarify the
 declarative semantics, further restrictions have to be placed on
 formulas to ensure coextensiveness with intuitionistic provability, a
 prelude to coextensiveness with uniform provability. By virtue of
 Theorem~\ref{thm:improrl}, one way to achieve this effect is to
 exclude the case involving implication from the syntax rule for
 $G$-formulas above. The language that results from this restriction
 is closely related to the Horn clause logic that  underlies the
 language Prolog: in particular, 
 it extends Horn clause logic as presented in \cite{MNPS91} by
 including universal quantification in goals and allowing $\bot$ to
 appear in goals and  program clauses. 

However, it is not necessary to exclude implications in goals even
when the chosen proof relation is classical provability.\footnote{This
  is not in contradiction to Theorem~\ref{thm:ctoicomp}. As noted
  already, the analysis in the
  theorem does not pay attention to the order in which rules are used
  and so is not fine-grained enough to provide a tight constraint on
  the syntax of formulas.} What the examples used in the proof of
 Theorem~\ref{thm:ctoicomp} show is that implications must not appear 
 negatively in program clauses or embedded within disjunctions or
 existential quantifications in goals. We can modify the definition of
 $G$- and $D$-formulas as follows to satisfy these requirements:
\[ \begin{array}{rcl}
 G & ::= & G' \sep D \supset G \sep G \land G \sep \allx{x}G \\
 G' & ::= & \top \sep \bot \sep A \sep G' \land G' \sep G' \lor G'
 \sep \allx{x}G' \sep \somex{x} G' \\ 
 D & ::= & \top \sep \bot \sep A \sep G'\supset D \sep D \land D \sep
 \sep \allx{x} D \end{array} \] 
Using Theorems~\ref{thm:improrl} and \ref{thm:inttounif} and the
 easy observations that (a)~$\sequent{\Gamma}{F_1 \land F_2}$ has a
 \Cbf-proof only if $\sequent{\Gamma}{F_1}$ and
 $\sequent{\Gamma}{F_2}$ also have \Cbf-proofs,
 (b)~$\sequent{\Gamma}{F_1 \supset F_2}$ has a \Cbf-proof only if
 $\sequent{F_1,\Gamma}{F_2}$ also has one, and
 (c)~$\sequent{\Gamma}{\allx{x}F}$ has a \Cbf-proof only if, for some
 constant $c$ not appearing in $\Gamma$ or $F$,
 $\sequent{\Gamma}{\subs{c}{x} F}$ also has one, it can be seen that these
 definitions in fact yield an abstract logic programming
 language. Moreover, this is the largest such language based on
 classical logic that also meets the mentioned requirements for
 programming. 

\subsection{Reduction to uniform provability}
The succedent formula can be used to direct the search for a uniform
proof for a sequent in a fairly deterministic fashion. 
By exploiting this fact, it is possible to define efficient proof
procedures for logical languages that have a derivability relation
that is coextensive with uniform provability. This idea has been used
previously relative to abstract logic programming languages; see, for
instance, \cite{miller91jlc,Nad92int}. 
Now, even in situations where the proof relation of interest deviates
from uniform provability, it may still be possible to utilize the
latter notion in structuring proof search. For instance, it may be
possible to modify the sequent whose derivability status is to be
verified in some predetermined and sound way to produce a new sequent
that has a derivation in the relevant sense just in case it has an
\Obf-proof. One approach of this kind that has been considered in the
past in conjunction with classical logic \cite{NL95lics,Nad96jlc}.   
In this approach, the attempt to prove a sequent of the form
$\sequent{\Gamma}{F}$ is transformed into an attempt to prove
$\sequent{F \supset \bot, \Gamma}{F}$ instead.
As we see below, the indicated augmentation of the antecedent
can be made implicit by being built into new inference rules. 
The virtue of the resulting derivation system is that it provides the
basis for a goal-directed proof procedure with the characteristic that
the attempt to prove the original goal is {\it restarted} with a modified
set of premises at certain points in the search
\cite{Gab85,GR93,LR91rob,Nad96jlc}. 

A crucial requirement in using this method is that the described
augmentation of the sequent reduce the question of classical
provability to that of uniform provability. Towards understanding the
applicability of this method, we wish to circumscribe the sequents for
which this reduction is actually achieved. We begin by observing that
the overall approach is actually sound:

\begin{lemma}\label{lem:equiv}
Let $\Gamma$ be a multiset of formulas and let $F$ be a formula. Then
$F \supset \bot, \Gamma \cprove F$ if and only if 
$\Gamma \cprove F$.  
\end{lemma}

\begin{proof}
This follows easily from the admissibility of the {\it Cut} inference
rules and the fact that $\sequent{}{(F \supset \bot) \lor F}$ has a
\Cbf-proof. 
\end{proof}

Since \Obf-proofs are \Ibf-proofs of a special form, a useful first
step towards the desired characterization is to understand 
when the augmentation of a sequent succeeds in reducing classical
provability to intuitionistic provability. 

\begin{theorem}\label{thm:ctoired}
Let $\Gamma$ be a multiset of formulas and let $F$ be a formula such
that $\sequent{\Gamma}{F}$ has a \Cbf-proof. Then there is an
\Ibf-proof for $\sequent{F \supset \bot, \Gamma}{F}$ if any one of the
following conditions holds relative to the \Cbf-proof of
$\sequent{\Gamma}{F}$: 
\begin{enumerate}
\item no \allr\ rule is used,  
\item no \impr\ and no \orl\ rule is used,
\item no \impr\ and no \alll\ rule is used, and 
\item no \impl, \orr\ and \somer\ rule is used.
\end{enumerate}
Further, for each way of violating all these conditions, there is a
sequent $\sequent{\Gamma}{F}$ with a violating \Cbf-proof such that
$\sequent{F \supset \bot,\Gamma}{F}$ does not have an \Ibf-proof.
\end{theorem}

\begin{proof}
Using the results in \cite{MO63}, it can be established that 
$\sequent{\Gamma}{F}$ has a \Cbf-proof in which no \allr\ rule is
used, then $\sequent{F \supset \bot,\Gamma}{F}$ has an
\Ibf-proof. This fact is also independently and explicitly established
in \cite{Nad96jlc}. Further, it is obvious that
$\sequent{F \supset \bot,\Gamma}{F}$ has an \Ibf-proof if
$\sequent{\Gamma}{F}$ has one. If any one of conditions 2-4 is true,
then, by virtue of Theorems~\ref{thm:improrl}, \ref{thm:impralll} and 
\ref{thm:implsomerorr}, $\sequent{\Gamma}{F}$ has an \Ibf-proof. Thus,
if any one of the listed conditions is true, then $\sequent{F\supset
  \bot,\Gamma}{F}$ must have an \Ibf-proof. 

It only remains to be shown that, corresponding to each way of
violating all the conditions, there is a sequent that has a \Cbf-proof
but whose augmented version does not have an \Ibf-proof. Clearly, we
need to consider only those situations in which a
\allr\ rule is used in the \Cbf-proof. Now, our analysis breaks up
into two parts, depending on whether or not a \impr\ rule appears in
the \Cbf-proof. Suppose, first, that it does not. Then the \Cbf-proof
must contain occurrences of both an \orl\ and a \alll\ rule and of
one of the \impl, \orr\ and \somer\ rules. The following sequents have
\Cbf-proofs respectively meeting each of these requirements:
\begin{tabbing}
\qquad\=\kill
\>$\sequent{\allx{x}\allx{y}(p(x) \lor q(y)), (\allx{x}p(x)) \supset
  (\allx{y}q(y))}{\allx{y}q(y)}$,\\
\>$\sequent{\allx{x}\allx{y}(p(x) \lor q(y))}{(\allx{x}p(x)) \lor
  (\allx{y}q(y))}$,\\
\>$\sequent{\allx{x}\allx{y}(r(x,a) \lor r(y,b))}{\somex{y}\allx{x}r(x,y)}$.
\end{tabbing}
We assume that $p$ and $q$ are unary predicate symbols, that $r$ is a
binary predicate symbol and that $a$ and $b$ are constants in these
sequents. Now, denoting the antecedent by $\Gamma$ and the formula in
the succedent by $F$ in each case, it can be seen that in none of
these cases does $\sequent{F \supset \bot, \Gamma}{F}$ have an
\Ibf-proof. 

To complete the argument, we consider the situation in which an \impr\
rule appears in the \Cbf-proof. In this case, one of the \impl, \orr\
and \somer\ must also appear in the \Cbf-proof. But then consider the
following sequents: 
\begin{tabbing}
\qquad\=\kill
\>$\sequent{\allx{x}((p(x) \supset \bot) \supset
  \bot)}{\allx{x}p(x)}$,\\
\>$\sequent{}{\allx{x}(p(x) \lor (p(x) \supset s))}$, and\\
\>$\sequent{}{\somex{y}\allx{x}(p(y) \supset p(x))}$. 
\end{tabbing}
In these sequents we assume additionally that $s$ is a proposition
symbol. Now, these sequents have \Cbf-proofs respectively meeting each
of the requirements. However, it can be easily seen that none of the
augmented versions of these sequents have an \Ibf-proof. 
\end{proof}

\ignore{Theorem~\ref{thm:ctoired} defines, in a sense, the limits of the
method under consideration for transforming classical provability into
uniform provability. We now show that this limit is actually
achieved. Part of this task has been carried out in
\cite{Nad96jlc}. In particular, it has been shown there that if no
\allr\ rule appears in the \Cbf-proof of a sequent, then the augmented
version of the sequent has an \Obf-proof. The remaining task, then, is
to show that if a knowledge of only the kinds of inference rules
used in a \Cbf-proof of a sequent assures us of the existence of an
\Ibf-proof for it, then the augmented version of the sequent has an
\Obf-proof. 
Towards this end, we find it convenient to modify our calculus for
constructing \Ibf-proofs. 
In particular, consider the following inference rules that are
parameterized by a specific formula $G$: }

It remains only to characterize the situations in which the
augmentation suffices to reduce intuitionistic provability to uniform
provability. Part of this task has already been performed in
\cite{Nad96jlc}. In particular, it has been shown there that if
$\sequent{G \supset \bot, \Gamma}{G}$ has an \Ibf-proof in which no
\allr\ rule is used, then this sequent also has an \Obf-proof. In
determining the other situations in which a similar property holds, 
we find it convenient to use a modified version of our calculus for
constructing \Ibf-proofs. 
Towards this end, we consider the following inference rules that are
parameterized by a specific formula $G$: 

\medskip

\begin{center}
\mbox{\inrulebn
         {\sequent{B,\Delta}{F}}
         {\sequent{D,\Delta}{G}}
         {\sequent{B \lor D, \Delta}{F}}
         {\mthorlg}}

\bigskip

\mbox{\inrulean
         {\sequent{\Delta}{G}}
         {\sequent{\Delta}{F}}
         {\mthresg}}
\end{center}

\medskip

\noindent We assume that $B$, $D$ and $F$ are schema variables for
formulas in these rules and that $\Delta$ denotes a multiset of
formulas. These rules are obviously derived ones relative to the
calculus for constructing \Ibf-proofs in the case that $\Delta$
contains the formula $G \supset \bot$. 
Moreover, every use that is made of the ``additional'' formula $G
\supset \bot$ in an \Ibf-proof of $\sequent{G\supset \bot, \Gamma}{G}$
can actually be transformed into a use of a \resg\ rule. 
Thus, in constructing an \Ibf-proof of a sequent of the form
$\sequent{G \supset \bot, \Gamma}{G}$, we may use these rules and also
make the augmentation of the antecedent implicit by strengthening
the proviso on the \somel\ and \allr\ rules to disallow the use of
constants appearing in $G$. We are actually interested in a calculus
that results from the above modifications and the {\it removal} of the
\orl\ rule. Let us refer to derivations constructed within this
calculus as \Ibfg-proofs. We then have the following observation.

\begin{lemma}\label{lem:orltoorlg} 
Let the sequent $\sequent{G \supset \bot, \Gamma}{G}$ have an
  \Ibf-proof in which no \alll\ or \impr\ rules are used. Then 
  there is an \Ibfg-proof for $\sequent{\Gamma}{G}$ in which no \alll\
  and \impr\ rules are used. 
\end{lemma}

\begin{proof} 
By an \Ibfgprime-proof let us mean a derivation that does
  not contain any \alll\ or \impr\ rules and that would be an
  \Ibfg-proof except for the fact that some number of \orl\ rules
  appear in it. From the premises of the lemma, it follows that
  $\sequent{\Gamma}{G}$ has an \Ibfgprime-proof. Thus, it suffices to
  show that an \Ibfgprime-proof of  
  $\sequent{\Gamma}{G}$ with some \orl\ rules in it can be transformed
  into one that does not contain any \orl\ rules. 
  We do this by an inductive argument based on the number of \orl\
  rules in the given \Ibfgprime-proof. We shall assume in this argument
  that this derivation satisfies two additional properties: (a)~the
  antecedent(s) of the upper sequent(s) of each left operational rule
  contains (contain) an occurrence of the principal formula of that rule and
  (b)~each \somel\ and \allr\ rule uses a distinct constant all of
  whose occurrences are restricted to the part of the derivation
  appearing above that rule. We may have to introduce some
  \contl\ rules into the original derivation to make 
  sure that the first requirement is satisfied and a consistent
  renaming of some constants suffices to ensure the second property.
  These `preprocessing' steps may be applied with impunity since they
  do not increase the number of \orl\ rules in the derivation and they
  also produce something that is itself an \Ibfgprime-proof of the
  same final sequent. 

An explicit argument is needed only in the case that at least one
\orl\ rule appears somewhere in the derivation. Suppose this happens
to be of the form  
\begin{center}
\mbox{\inrulewjan
          {\sequent{B,\Sigma}{F} \qquad\qquad \sequent{D,\Sigma}{F}}
          {\sequent{B \lor D, \Sigma}{F}}}
\end{center}

\noindent We may replace this with an \orlg\ rule, thereby reducing
the number of occurrences of \orl\ rules, provided we can produce an
\Ibfg-proof for $\sequent{D, \Sigma}{G}$. It suffices, for this
purpose, to exhibit an \Ibfgprime-proof for $\sequent{D, \Sigma}{G}$
with fewer \orl\ rules in it than in the given derivation for
$\sequent{\Gamma}{G}$. Such a derivation can be constructed based on
the one for $\sequent{\Gamma}{G}$ by retaining unchanged the
portion of the latter derivation above the sequent $\sequent{D,
\Sigma}{F}$ and by transforming the portion below this sequent
as follows: 
\begin{enumerate}
\item Replacing all left rules that are not \orlg\ rules above whose right
upper sequent $\sequent{D, \Sigma}{F}$ appears by the sequent
$\sequent{D,\Sigma}{F'}$ where $F'$ is the succedent of the upper and
lower sequents of this rule; applications of this transformation to
a sequence of such rules will result in replacement by a single
sequent.

\item Erasing the portion of the derivation up to and including the left
  upper sequent of all remaining \orlg\ rules and renaming these to
  \resg\ rules.

\item Replacing each right rule with an instance of the same schema
  but with $D, \Sigma$ as the antecedent of the upper and lower
  sequents.

\item Replacing the derivation above the upper sequent of an \andr\
  rule that is different from the one above which the sequent
  $\sequent{D, \Sigma}{F}$ appears by one that uses the same rule
  schemata but with suitably modified antecedents. 
\end{enumerate}
Clearly, this construction eliminates at least one \orl\ rule from the
given \Ibfgprime-proof. However, some care is needed in ascertaining that it
yields something that is indeed an \Ibfgprime-proof. First, each \allr\
rule below $\sequent{D, \Sigma}{F}$ in the new `derivation' uses the
same constant as is used in the derivation of $\sequent{\Gamma}{G}$
and we must verify that this is acceptable. We see this to be the case
by observing that this constant cannot appear in $D, \Sigma$ since the
given derivation does not contain occurrences of either the \alll\ or
the \impr\ rules. Second, it must be possible to construct the
derivation above the other upper sequent of an \andr\ rule as
described; in particular, all the \allr\ and left rules needed in this
construction must be legitimate ones. Our assumptions concerning the
constants used in \somel\ and \allr\ rules and the relationship
between the antecedents of the upper and lower sequents of each left
operational rule in the given \Ibfgprime-proof ensure that this is the
case. 
\end{proof}

We now relativize the notion of a uniform proof to our modified
calculus. 
In particular, let an \Obfg-proof be an \Ibfg-proof with the following
characteristic: if there is a sequent in this proof whose succedent
contains a non-atomic formula, then that sequent occurs as the
lower sequent of an inference rule that introduces the top-level
logical symbol of that formula. 
The following may then be observed:

\begin{lemma}\label{lem:secondmodified}
If $\sequent{\Gamma}{G}$ has a \Ibfg-proof in which no \alll\ or
\impr\ rules appear, then it has an \Obfg-proof.  
\end{lemma}

\begin{proof}
By the {\it nonuniformity measure} of a left rule in an \Ibfg-proof
let us mean the count of right rules pertaining to logical
symbols in the succedent of the lower sequent of the left rule that
appear {\it above} the left rule in the derivation. 
Further, let the nonuniformity measure of the \Ibfg-proof itself be 
defined to be the sum of the nonuniformity measures of the left
operational rules contained in it. 
Now, let us refer to an \Ibfg-proof in which no \alll\ or \impr\ rules
appear as an \Ibfgprime-proof.  
We claim then that if $\sequent{\Gamma}{G}$ has an \Ibfgprime-proof,
then it has one whose nonuniformity measure is $0$. We prove this
claim by induction on the measure. We shall assume
in our argument that the given derivation satisfies two additional
properties: (a)~the antecedent(s) of the upper sequent(s) of each left
operational rule contains (contain) an occurrence of the principal
formula of that rule and (b)~each \somel\ and \allr\ rule uses a distinct
constant all of whose occurrences are restricted to the part of the
derivation appearing above that rule. 
We may have to apply the preprocessing steps discussed in the proof of
Lemma~\ref{lem:orltoorlg} to ensure that these requirements are met,
but we can do this without changing the nonuniformity measure of the
derivation. 

In order to establish the claim, it is sufficient to show that if
$\sequent{\Gamma}{G}$ has an \Ibfgprime-proof with nonzero
nonuniformity measure, then it has one with a smaller such measure.
From the assumption it follows that the given derivation contains
a left operational rule with right operational rules pertaining to the
succedent of its lower sequent appearing above it. We focus on a 
left rule that is the {\it first} along some path in the derivation
to have this characteristic. It is easily seen that a \contl\
rule can be moved above any right rule in an \Ibfg-proof. Thus, we may
assume that the left operational rule of interest appears {\it
  immediately after} the relevant right rule in the given
\Ibfgprime-proof. Our objective, now, is to show that these two rules
can be reordered in a way that decreases the nonuniformity measure of
the overall derivation. 

A simple transformation can be used to achieve this effect when
the left rule is not a \somel\ or the right rule 
is not a \somer. We illustrate this by considering one particular
case: that when the left rule is an \impl\ and the right rule is an
\andr. In this case, the subderivation at the end has the following
structure:
\begin{center}
\mbox{\inrulebn{\sequent{\Delta}{B}}
               {\rinrulebn{\sequent{D,\Delta}{F_1}}
                         {\sequent{D,\Delta}{F_2}}
                         {\sequent{D,\Delta}{F_1 \land F_2}}
                         {\mthandr}}
               {\sequent{B \supset D, \Delta}{F_1 \land F_2}}
               {\mthimpl}}
\end{center}
\noindent By assumption, the nonuniformity measure of the derivation
of $\sequent{\Delta}{B}$ is $0$.
We may reuse the \Ibfgprime-proofs of $\sequent{\Delta}{B}$,
$\sequent{D,\Delta}{F_1}$ and $\sequent{D,\Delta}{F_2}$ to produce an
alternative subderivation of $\sequent{B \supset D, \Delta}{F_1 \land
  F_2}$ that has the structure
\begin{center}
\mbox{\inrulebn{\linrulebn{\sequent{\Delta}{B}}
                         {\sequent{D,\Delta}{F_1}}
                         {\sequent{B \supset D, \Delta}{F_1}}
                         {\mthimpl}}
               {\rinrulebn{\sequent{\Delta}{B}}
                         {\sequent{D,\Delta}{F_2}}
                         {\sequent{B \supset D, \Delta}{F_2}}
                         {\mthimpl}}
               {\sequent{B \supset D, \Delta}{F_1 \land F_2}}
               {\mthandr}}
\end{center}
\noindent at the end. The nonuniformity measure of the new
subderivation is obviously less than that of the earlier
one, and it also does not have any new occurrences of right rules that
could increase the nonuniformity measure of left operational rules
appearing later in the derivation. Thus, the desired effect is
achieved by this transformation.

For the only remaining case, let us suppose that it occurs in a 
subderivation that has the structure
\begin{center}
\mbox{\inrulean{\inrulean
                  {\sequent{\subs{c}{x}B, \Delta}{\subs{t}{y}D}}
                  {\sequent{\subs{c}{x}B, \Delta}{\somex{y}D}}
                  {\mthsomer}}
               {\sequent{\somex{x}B, \Delta}{\somex{y}D}}
               {\mthsomel}}
\end{center}
\noindent at the end. Now, it can be shown that
$\sequent{\subs{c}{x}B,\Delta}{G}$ has an \Ibfgprime-proof of smaller
nonuniformity measure than that of the one for $\sequent{\Gamma}{G}$; 
as in the proof of Lemma~\ref{lem:orltoorlg}, we construct such a
derivation essentially by mimicking the structure of the given
\Ibfgprime-proof of $\sequent{\Gamma}{G}$ and note that at least one
occurrence of a \somel\ rule---the one shown above---that makes a
nonzero contribution to the nonuniformity measure is eliminated in the
process. From the induction hypothesis, it follows then that 
$\sequent{\subs{c}{x}B,\Delta}{G}$ has an \Ibfgprime-proof of zero
nonuniformity measure. The proviso on a \somel\ rule ensures that $c$
does not occur in $B$, $\Delta$ or $G$. Given this, we may also 
assume that $c$ does not also appear in $t$, for, if it does, we
simply rename it to a new constant $c'$ that satisfies this additional
requirement and use $\sequent{\subs{c'}{x}B,\Delta}{G}$ and its
corresponding derivation in the rest of the argument. Using the known
derivation for 
$\sequent{\subs{c}{x}B,\Delta}{G}$, we may restructure the
\Ibfgprime-proof for $\sequent{\somex{x}B, \Delta}{\somex{y}D}$ so
that it has the 
form
\begin{center}
\mbox{\inrulean{\inrulean
                  {\inrulean
                         {\sequent{\subs{c}{x}B, \Delta}{G}}
                         {\sequent{\subs{c}{x}B,
                             \Delta}{\subs{t}{y}D}}
                         {\mthresg}}
                  {\sequent{\somex{x}B, \Delta}{\subs{t}{y}D}}
                  {\mthsomel}}
               {\sequent{\somex{x}B, \Delta}{\somex{y}D}}
               {\mthsomer}}
\end{center}
\noindent at the end. This derivation obviously has a
nonuniformity measure less than that of the earlier one and using it
instead also decreases the nonuniformity measure of the overall
derivation. 

We have thus shown that $\sequent{\Gamma}{G}$ has an \Ibfgprime-proof,
and, hence, an \Ibfg-proof, of zero nonuniformity measure. By moving
\contl\ rules above any 
immediately preceding right rules in this derivation, we obtain a
structure that would be an \Obfg-proof if an additional property
holds: the succedent of the lower sequent of every \botr\ and \resg\
rule is an atomic formula. 
This may not be true at the outset, but a simple transformation
process ensures that it eventually is. To illustrate this process, suppose that
there is a \resg\ rule in the derivation whose lower sequent has the
formula $F_1 \land F_2$ as its succedent. Now, there must be a last 
sequent following this one in the derivation that has the same formula
as its succedent. Suppose this sequent is $\sequent{\Delta}{F_1 \land
  F_2}$. By imitating the derivation of this sequent, we obtain
\Ibfg-proofs for $\sequent{\Delta}{F_1}$ and
$\sequent{\Delta}{F_2}$. Further, using these \Ibfg-proofs, we may
replace the derivation of $\sequent{\Delta}{F_1 \land F_2}$ by one
that has the structure
\begin{center}
\mbox{\inrulebn{\sequent{\Delta}{F_1}}
               {\sequent{\Delta}{F_2}}
               {\sequent{\Delta}{F_1 \land F_2}}
               {\mthandr}}
\end{center}
at the end without changing the nonuniformity measure of the overall
\Ibfg-proof. The virtue of this transformation is that the \resg\ rule
in the original derivation is replaced by ones whose lower sequent
have formulas with fewer logical symbols in their
succedents. In a more detailed presentation, we associate with each
\Ibfg-proof of zero nonuniformity measure a multiset of numbers that
count the logical symbols in the formulas that appear as the
succedents of the lower sequents of \botr\ and \resg\ rules used in
the derivation. We then use the above form of argument in an induction
over the multiset ordering induced by the usual ordering on natural numbers
\cite{Dershowitz82} to show that $\sequent{\Gamma}{G}$ has an
\Ibfg-proof of zero nonuniformity measure and in which the succedent
of the lower sequent of every \botr\ and \resg\ rule is atomic. 
\end{proof}

The following theorem states the desired relationship between
intuitionistic and uniform provability.

\begin{theorem}\label{thm:itoored}
Suppose that there is an \Ibf-proof for a sequent of the form
$\sequent{G \supset \bot, \Gamma}{G}$ satisfying one of the following
restrictions on rule usage:
\begin{enumerate}
\item No \allr\ rule is used.

\item No \orr\ or no \orl\ rule is used and, in addition, either no
  \somer\ rule is used or no \orl\ and \somel\ rules are used.

\item no \alll\ and \impr\ rules are used.
\end{enumerate}
Then there is an \Obf-proof for the same sequent. Furthermore, this
characterization is complete in the following sense: there is a
sequent of the required form that has an \Ibf-proof but no \Obf-proof
corresponding to each way of violating all the restrictions on
inference rule usage. 
\end{theorem}

\begin{proof} The sufficiency of the first restriction on inference
  rule usage is shown in \cite{Nad96jlc} and that of the second
  restriction follows immediately from
  Theorem~\ref{thm:inttounif}. The sufficiency of the third
  restriction is a consequence of Lemmas~\ref{lem:orltoorlg} and
  \ref{lem:secondmodified} and the observation that an \Obfg-proof for
  $\sequent{\Gamma}{G}$ can be translated into a uniform proof for
  $\sequent{G \supset \bot, \Gamma}{G}$.  

We now show the completeness of the characterization in the sense
claimed. To begin with, the only situations we need to consider are
those in which a \allr\ 
rule is used in the \Ibf-proof. Now, we may partition these situations
based on whether an \orr\ or a \somer\ rule has been used. Considering
the former possibility first, we note that in these situations an 
\orl\ rule and one of the \alll\ and \impr\ rules must also have been
used. The following sequents have \Ibf-proofs respectively satisfying these
requirements on rule usage:
\begin{tabbing}
\qquad\=\qquad\qquad\qquad\qquad\qquad\=\kill
\>$\sequent{(\allx{y}(r(b,y) \lor r(a,y))) \supset \bot,
  \allx{y}(r(a,y) \lor r(b,y))}{\allx{y}(r(b,y) \lor r(a,y))}$, and \\
\>$(\allx{y}((r(b,y) \lor r(a,y)) \supset (r(a,y) \lor
  r(b,y)))) \supset \bot \ \longrightarrow$\\
\>\>$\allx{y}((r(b,y) \lor r(a,y)) \supset (r(a,y) \lor r(b,y)))$;
\end{tabbing}
\noindent we assume that $r$ is a binary predicate symbol and $a$ and
$b$ are constants in these sequents. It is easily seen that neither of
these sequents has an \Obf-proof, as is required. 

To finish the proof, we have to consider those situations in which the
violation of the restrictions arises from the use of a \somer\
rule. In these cases, one of the \orl\ and \somel\ rules and also one
of the \alll\ and \impr\ rules must also have been used. We list four
sequents of the required form that have \Ibf-proofs respectively
satisfying these requirements:
\begin{tabbing}
\qquad\=\kill
\>$\sequent{(\allx{y}\somex{x}r(x,y)) \supset \bot, \allx{y}(r(a,y)
  \lor r(b,y))}{\allx{y}\somex{x}r(x,y)}$,\\
\>$\sequent{(\allx{y}((r(a,y) \lor r(b,y)) \supset \somex{x}r(x,y)))
  \supset \bot}{\allx{y}((r(a,y) \lor r(b,y)) \supset
  \somex{x}r(x,y))}$,\\
\>$\sequent{(\allx{y}\somex{x}r(x,y)) \supset \bot,
  \allx{y}\somex{x}r(x,y)}{\allx{y}\somex{x}r(x,y)}$, and \\
\>$\sequent{(\allx{y}((\somex{x}r(x,y)) \supset \somex{x}r(x,y)))
  \supset \bot}{\allx{y}((\somex{x}r(x,y)) \supset \somex{x}r(x,y))}$.
\end{tabbing}
\noindent Once again, it can be verified that none of these sequents
has an \Obf-proof.
\end{proof}

Combining Theorems~\ref{thm:ctoired} and \ref{thm:itoored}, we see
that if $\sequent{\Gamma}{G}$ has a \Cbf-proof satisfying one of the
following restrictions on rule usage, then there must also be an
\Obf-proof for $\sequent{G\supset \bot,\Gamma}{G}$: (i)~no \allr\ rule
is used, (ii)~no \impl, \orr\ and \somer\ rule is used, and (iii)~no
\impr\ and \alll\ rule is used. These restrictions can be recast in an
obvious manner into ones on the syntax of formulas in the sequent for
which a derivation is to be constructed and are, in fact, more useful
in this form. Of all these conditions, the one that is most easily
ensured in practice is that there be no universal quantifiers
occurring negatively in the antecedent and positively in the
succedent---any sequent can be transformed in one that is equivalent
from the perspective of classical provability and that satisfies this
additional property through the use of Herbrand functions
\cite{Shankar92}. A proof procedure based on these observations is
described in \cite{Nad96jlc} and connections with other previously
presented procedures is also discussed there. 

We observe, finally, that the results of this section are also
relevant from the perspective of structuring proof search in
intuitionistic logic. In particular, the augmentation of sequents is
sound with respect to intuitionistic provability whenever the
structure of the sequent ensures a coincidence with classical
provability. Such an augmentation may then be used to obtain a
reduction to uniform provability. One interesting situation in which
this approach may be utilized is that when implications and universal
quantifications do not appear positively in the succedent and
negatively in the antecedent of a sequent. This situation epitomizes
disjunctive logic programming and is discussed in more detail in
\cite{NL95lics}. 

\section{Conclusion}\label{sec:conc}
We have explored the interrelationships between the notions of
classical, intuitionistic and uniform provability in this paper. 
We have also examined the relevance of our results to proof search in
classical and intuitionistic logic and to identifying logic
programming languages. We believe there are other applications to our
observations as well, especially to our characterization of the
correspondence between classical and intuitionistic
provability. Another matter that is only partially studied here and
that is worthy of further consideration is the usefulness of uniform
provability in designing proof procedures for intuitionistic logic. 

\section{Acknowledgements}\label{sec:ack}

We are grateful for comments received from the anonymous
reviewers of this paper; the careful reading and detailed suggestions
of one of these reviewers have led to several improvements in
presentation. We also acknowledge helpful comments provided by Michael
J. O'Donnell. This work has been supported in part by NSF Grants
CCR-92-08465 and CCR-9596119.

\end{document}